\documentclass[twocolumn]{aastex62}
\usepackage{apjfonts}
\usepackage{here}
\submitjournal{ApJ}

\shorttitle{Spatially resolved spectroscopy of thermal X-rays in IC~443 with {\it XMM-Newton}}
\shortauthors{Okon et al.}
\begin{document}

\title{Investigation of the Physical Origin of Overionized Recombining Plasma in the Supernova Remnant IC~443 with {\it XMM-Newton}}

\correspondingauthor{Hiromichi Okon}
\email{hiromichi.okon@cfa.harvard.edu}

%\author{Hiromichi Okon}
%\affil{Department of Physics, Kyoto University, Kitashirakawa Oiwake-cho, Sakyo, Kyoto 606-8502, Japan}
\author{Hiromichi Okon}
\affil{Harvard-Smithsonian Center for Astrophysics, 60 Garden Street, Cambridge, MA 02138, USA}

\author{Takaaki Tanaka}
\affil{Department of Physics, Konan University, 8-9-1 Okamoto, Higashinada, Kobe, Hyogo 658-8501, Japan}
	
\author{Hiroyuki Uchida}
\affil{Department of Physics, Kyoto University, Kitashirakawa Oiwake-cho, Sakyo, Kyoto 606-8502, Japan}	

\author{Takeshi Go Tsuru}
\affil{Department of Physics, Kyoto University, Kitashirakawa Oiwake-cho, Sakyo, Kyoto 606-8502, Japan}

\author{Masumichi Seta}
\affil{Department of Physics and Astronomy,School of Science, Kwansei Gakuin University 2-1 Gakuen, Sanda, Hyogo 669-1337, Japan}

\author{Takuma Kokusho}
\affil{Graduate School of Science, Nagoya University, Furo-cho, Chikusa, Nagoya 464-8602, Japan}

\author{Randall K. Smith}
\affil{Harvard-Smithsonian Center for Astrophysics, 60 Garden Street, Cambridge, MA 02138, USA}
%
%\author{}
%\affil{Department of Physics, Kyoto University, Kitashirakawa Oiwake-cho, Sakyo, Kyoto, Kyoto 606-8502, Japan}
%
%\author{}
%\affil{Department of Physics, Kyoto University, Kitashirakawa Oiwake-cho, Sakyo, Kyoto, Kyoto 606-8502, Japan}
%



\begin{abstract}

The physical origin of the overionized recombining plasmas (RPs) in supernova remnants (SNRs) has been attracting attention because its understanding provides new insight into SNR evolution.
However, the process of the overionization, although it has been discussed in some RP-SNRs, is not yet fully understood.
Here we report on spatially resolved spectroscopy of X-ray emission from IC~443 with {\it XMM-Newton}.
We find that RPs in regions interacting with dense molecular clouds tend to have lower electron temperature and lower recombination timescale.
These tendencies indicate that RPs in these regions are cooler and more strongly overionized, which is naturally interpreted as a result of rapid cooling by the molecular clouds via thermal conduction.
Our result on IC~443 is similar to that on W44 showing evidence for thermal conduction as the origin of RPs at least in older remnants.
We suggest that evaporation of clumpy gas embedded in a hot plasma rapidly cools the plasma as was also found in the W44 case.
We also discuss if ionization by protons accelerated in IC~443 is responsible for RPs.
Based on the energetics of particle acceleration, we conclude that the proton bombardment is unlikely to explain the observed properties of RPs.

\end{abstract}

%% keywords and the roles for their use.


\section{Introduction} 
%\linenumbers

X-ray spectroscopy of supernova remnants (SNRs) enables us to investigate the thermal properties of SNR plasmas, providing information on their evolution history.
Recent X-ray observations unveiled the presence of plasmas in a recombination-dominant state in a dozen SNRs \cite[e.g.,][]{Ozawa2009}, including the target of the present work, IC~443 \citep{Yamaguchi2009}.
Such plasmas have a higher ionization degree than that expected in collisional ionization equilibrium (CIE), and thus often been called overionized recombining plasmas (RPs).
RPs were not anticipated in the previously accepted scenario that SNR plasmas are collisonally ionized until they reach equilibrium.
The physical origin of the overionization has been attracting attention because it reveals important processes not considered in the standard scenario \cite[e.g.,][]{Zhang2019}.

The formation process of RPs is thought to be closely related to interaction between the SNRs and ambient clouds, and two distinct scenarios have been mainly considered.
%One one scenario is called the thermal conduction scenario and is based on an idea originally proposed by \cite{Kawasaki2002,Kawasaki2005}.
One model is the thermal conduction scenario, which is based on an idea originally proposed by \cite{Kawasaki2002,Kawasaki2005}.
Some authors, e.g., \cite{Matsumura2017a}, \cite{Okon2018}, and \cite{Katsuragawa2018}, who analyzed {\it Suzaku} data of G166$+$4.3, W28, and CTB~1, respectively, claimed that thermal conduction by ambient clouds cools their X-ray plasmas and makes the overionization.
%\cite{Sano2021} reached the same conclusion using ALMA data of W49B.
%\cite{Sano2021} provided the supportive result for this cooling process by ALMA data.
%\cite{Sano2021} provided another reinforcing the cooling process with ALMA data. 
%\cite{Sano2021} claimed the process can contribute to the overionization with ALMA data
%The process can contribute to the overionization as reported by \cite{Sano2021} who analyzed ALMA data.
%\cite{Sano2021} claimed the process contribute to the overionization in W49B using ALMA data.
Carrying out spatially resolved spectroscopy of X-ray emission of W44 with {\it XMM-Newton}, \cite{Okon2020} investigated spatial variations of the electron temperature and the overionization degree of the RP, and presented clear evidence for the thermal conduction scenario.

Another plausible scenario is rarefaction, predicted by \cite{Itoh1989} and \cite{Shimizu2012}, where rapid adiabatic expansion is responsible for the overionization.
%The observational results supporting this scenario are indeed reported by some authors, e.g., \cite{Miceli2010} and \cite{Lopez2013}.
%Observational results on W49B preferring this scenario are reported by some authors, e.g., \cite{Miceli2010}, \cite{Lopez2013}, and \cite{Ashford2020} although they are incompatible with on the work based recent ALMA data \citep{Sano2021}.
%Observational results on W49B preferring this scenario are reported by some authors, e.g., \cite{Miceli2010}, \cite{Lopez2013}, and \cite{Ashford2020} although their claims are incompatible with that based on recent ALMA data \citep{Sano2021}.
Observational results on W49B preferring this scenario are reported by some authors, e.g., \cite{Miceli2010}, \cite{Lopez2013}, and \cite{Ashford2020}, although they are questioned by \cite{Sano2021} based on recent ALMA data.
%Observational results on W49B preferring this scenario are reported by some authors, e.g., \cite{Miceli2010} and \cite{Lopez2013} .
%Observational results on W49B preferring this scenario are reported by some authors, e.g., \cite{Miceli2010}, \cite{Lopez2013} and \cite{Ashford2020}.
\cite{Yamaguchi2018}, who performed spatially resolved spectroscopy of the remnant with {\it NuSTAR}, revealed obvious spatial variations of RP parameters, which are naturally explained by the rarefaction scenario.
%Although at least above mechanisms would be responsible for RPs and the dominant channel may change from adiabatic expansion to thermal conduction depending on their evolution stage \citep[e.g.,][]{Zhang2019,Okon2020}, the number of SNRs to which such analysis demonstrated by \cite{Yamaguchi2018} and \cite{Okon2020} was applied is still limited so that the formation process of RPs in SNRs is not comprehensively understood.
Although both of mechanisms could be responsible for RPs, the number of SNRs with such spatial-spectral analysis as demonstrated by \cite{Yamaguchi2018} and \cite{Okon2020} has been applied is still limited so that the formation process of RPs in SNRs is not comprehensively understood.

IC~443 (a.k.a., G189.1$+$3.0) has been regarded as one of the most important SNRs for studies of RPs.
% since \cite{Kawasaki2005} discussed the possibility of the RPs with {\it ASCA} data.
IC~443 is a Galactic core-collapse SNR at a distance of $\sim$1.5~kpc \citep{Welsh2003} toward the Galactic anticenter.
%The age of IC~443 is estimated to be $\sim$ 20--30 kyr \citep{Olbert2001,Lee2008}.
%\cite{Kawasaki2002} measured the Ly$\alpha$ to He$\alpha$ line intensity ratio of Si and S
Possible signatures of RPs were first suggested by \cite{Kawasaki2002} by measuring the He$\alpha$/Ly$\alpha$ ratio of Si and S with {\it ASCA}.
\cite{Yamaguchi2009} discovered radiative recombining continua (RRCs) of Si and S in {\it Suzaku} data, and showed first unequivocal evidence of an RP in SNRs, along with the W49B case by \cite{Ozawa2009}.
Subsequent {\it Suzaku} observations revealed that heavier ions such as Ca and Fe are also in the overionized state \citep{Ohnishi2014}.
IC~443 is known to be interacting with dense clouds through detections of $\rm ^{12}CO$ lines \citep[e.g.,][]{Denoyer1979,Seta1998} and $\rm H_2$ emissions \citep[e.g.,][]{Rho2001}.
Based on spectroscopic analyses of the X-ray data, \cite{Matsumura2017b} and \cite{Greco2018} claimed that thermal conduction and adiabatic expansion create the RP, respectively.
\cite{Hirayama2019} and \cite{Yamauchi2021} proposed a new scenario completely different from above two scenarios.
In this model, protons accelerated in SNRs enhance ionization of ions in plasmas, resulting in the overionization.
All of the authors' claims contradict each other and the physical origin of the RP in IC~443 is thus still been under active debate.

%Here we report on results from a spatially resolved analysis of {\it XMM-Newton} data of IC~443, toward the comprehensive understanding of the physical origin of the RP.
Here we report on results from a spatially resolved analysis of {\it XMM-Newton} data of IC~443, aiming to reveal the physical origin of the RP.
Throughout the paper, errors are quoted at 90\% confidence levels in the tables and text. 
Error bars shown in figures correspond to 1$\sigma$ confidence levels.

\section{Observations and Data Reduction}

\begin{figure*}[ht]
\begin{center}
\vspace{-0mm}
\includegraphics[width=16cm]{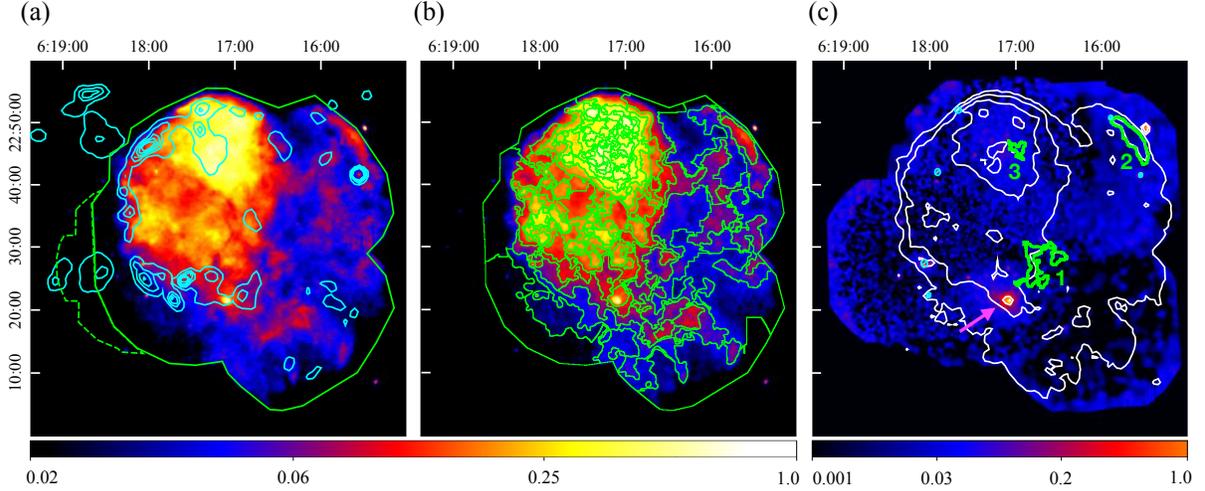} 
\end{center}
\vspace{-2mm}
\caption{
MOS$+$pn images of IC~443 in the energy band of (a)(b) 0.5--4.0~keV and (c) 4.0--8.0~keV after NXB subtraction and correction for the vignetting effect. 
The coordinate refers to the J2000.0 epoch. 
The cyan contours in panel (a) indicate a radio continuum image at 1.4 GHz taken with the the NRAO VLA Sky Survey.
The source and background spectra were extracted from the regions enclosed by the solid and dashed lines in panel (a), respectively.
The source region was divided into 110 sub-regions as shown in panel (b). 
The regions enclosed by the green lines in panel (c) are the three representative subregions whose spectra are plotted in Figure~\ref{fig:all_spectrum}. 
The white contours show the 0.5--4.0 keV X-ray image. 
The magenta arrow indicates the location of the PWN 1SAX J0617.1$+$2221.
The cyan ellipses are regions excluded in the spectral analysis to remove bright point sources. 
}
\label{fig:Xray_image}
\end{figure*}

\begin{deluxetable*}{lcccc}
\tablecaption{Observation log. \label{tab:obs_id}}
\tablecolumns{5}
%\tablenum{2}
\tablewidth{0pt}
\tablehead{
\colhead{Target} &
\colhead{Obs. ID} &
\colhead{Obs. Date} &
\colhead{(R.A., Dec.)\tablenotemark{a}} &
\colhead{Effective Exposure}
}
\startdata
     	%PSR~J1856+0113 & 0112201101 &2003 March 27 & ($18^{\rm h} 56^{\rm m} 11\fs00,~+01\degr 13\arcmin 21\farcs0$) & 2~ks \\%2.0~ks \\
	IC~443 & 0114100101 & 2000 September 26 & ($6^{\rm h} 17^{\rm m} 27\fs99,~+22\degr 41\arcmin 44\farcs0$) & 11~ks\\
	IC~443 & 0114100201 & 2000 September 25 & ($6^{\rm h} 16^{\rm m} 14\fs99,~+22\degr 41\arcmin 60\farcs0$) & 5~ks\\
	IC~443 & 0114100301 & 2000 September 27 & ($6^{\rm h} 17^{\rm m} 27\fs99,~+22\degr 25\arcmin 14\farcs0$) & 22~ks\\
	IC~443 & 0114100401 & 2000 September 28 & ($6^{\rm h} 16^{\rm m} 14\fs99,~+22\degr 18\arcmin 02\farcs5$) & 23~ks\\
	IC~443 & 0114100501 & 2000 September 25 & ($6^{\rm h} 16^{\rm m} 14\fs99,~+22\degr 41\arcmin 60\farcs0$) & 15~ks\\
	IC~443 & 0114100601 & 2000 September 27 & ($6^{\rm h} 17^{\rm m} 27\fs99,~+22\degr 25\arcmin 14\farcs0$) & 4~ks\\
	IC~443 & 0301960101 & 2006 March 30 & ($6^{\rm h} 18^{\rm m} 04\fs99,~+22\degr 27\arcmin 33\farcs0$) & 46~ks\\
	IC~443 & 0600110101 & 2010 March 17 & ($6^{\rm h} 17^{\rm m} 06\fs00,~+22\degr 43\arcmin 24\farcs0$) & 26~ks\\
\enddata
\label{tab:use_data}
\tablenotetext{a}{Equinox in J2000.0.}
\end{deluxetable*}

IC~443 was observed several times from 2003 to 2013 with {\it XMM-Newton}.
Table~\ref{tab:use_data} summarizes the details of the observations.
Once all observations are combined, the entire region of IC~443 is  completely covered by the field of view of {\it XMM-Newton}.
In what follows, we analyze data obtained with the European Photon Imaging Camera (EPIC), which consists of two MOS cameras \citep{Turner2001}, and one pn CCD camera \citep{Struder2001}.

We reduced the data with the Science Analysis System software version 16.0.0, following the cookbook for analysis procedures of extended sources\footnote{ftp://xmm.esac.esa.int/pub/xmm-esas/xmm-esas.pdf}.
We used the calibration database version 3.12 released in 2019\footnote{https://xmmweb.esac.esa.int/docs/documents/CAL-TN-0018.pdf}.
%We estimated the non- X-ray background (NXB) with {\tt mos-back}.
We generated the redistribution matrix files and the ancillary response files. 
%In the image analysis, we merged MOS1, MOS2, and pn data of each observation for better photon statistics, whereas,
%in the spectral analysis, we only used the MOS data because of their lower detector background level than the pn data.

\section{Analysis and Results} \label{sec:floats}
\subsection{Imaging Analysis}

%Figure~\ref{fig:Xray_image} shows vignetting- and exposure-corrected images of IC~443 after NXB subtraction.
Figure~\ref{fig:Xray_image} shows vignetting- and exposure-corrected images of IC~443.
Overlaid in Figure~\ref{fig:Xray_image}(a) is the source region for spectral analysis. 
%The source region is overlaid in Figure~\ref{fig:Xray_image}(a).
%To perform spatially resolved spectroscopic analysis, we applied the same method as that used in the image analysis of W44 \citep{Okon2020}.
We perform spatially resolved spectroscopy using the same method as \cite{Okon2020}.
We applied the contour-binning algorithm \citep{Sanders2006} to the 0.5--4.0 keV image, and divided the source region into 110 subregions in Figure~\ref{fig:Xray_image}(b). 
The algorithm generates subregions following the structure of the surface brightness so that each subregion has almost the same signal-to-noise ratio.
One can see the pulsar wind nebula (PWN) 1SAX J0617.1$+$2221, its pulsar (PSR), and some bright point sources in the 4.0--8.0 keV image, which are reported by \cite{Bocchino2003}.
For the following spectral analysis, we manually excluded the identified point sources whereas we modeled emissions from the PWN and PSR according to the work by \cite{Bocchino2003}.

\subsection{Background Estimation}\label{background}

For background estimation, we used spectra extracted from the off-source region in Figure \ref{fig:Xray_image}(a).
We subtracted the non-Xray background (NXB) estimated with {\tt mos-back} and {\tt pn-back} from the spectra, and fitted them with a model consisting of the X-ray background model \citep{Masui2009}, and neutral Al and Si K$\alpha$ lines of instrumental background, which are not included in the NXB spectra \citep{Lumb2002}.
%After subtracting the non-Xray background (NXB) estimated with {\tt mos-back} from the spectra, we fitted them with a model consisting of the X-ray background model \citep{Masui2009}, and neutral Al and Si K$\alpha$ lines of instrumental background, which are not included in the NXB spectra \citep{Lumb2002}.
The X-ray background model consists of  the cosmic X-ray background (CXB), the local hot bubble (LHB), and two thermal components for the Galactic halo (GH$_{\rm cold}$ and GH$_{\rm hot}$).
%We fitted the spectra with a model which consists of the cosmic X-ray background (CXB), the local hot bubble (LHB), two thermal components for the Galactic halo (GH$_{\rm cold}$ and GH$_{\rm hot}$), and neutral Al and Si K$\alpha$ lines of instrumental background not included in the estimated NXB spectra \citep{Lumb2002}.
The photon index of the CXB component was fixed to 1.4 given by \cite{Kushino2002} whereas the normalization and the column density ($N_{\rm H}^{\rm CXB}$) for the total Galactic absorption were allowed to vary.
We used the Tuebingen-Boulder interstellar medium (ISM) absorption model \cite[TBabs;][]{Wilms2000} with the solar abundances \citep{Wilms2000} for the interstellar absorption.
%For estimation of the LHB, GH$_{\rm cold}$, and GH$_{\rm cold}$, we employed the models of \cite{Masui2009}.
Most of the parameters of the LHB, GH$_{\rm hot}$, and GH$_{\rm cold}$ models were fixed to values given by \cite{Masui2009}.
The electron temperature $kT_e$ of GH$_{\rm hot}$, and the normalization of each component were left free.
The normalization of  the Al and Si K$\alpha$ lines were allowed to vary because the line intensities are known to have location-to-location variations on the detector plane \citep{Kuntz2008}.
The best-fit parameters are summarized in Table~\ref{tab:bkg_model}.
In the subsequent spectral analyses, we used the best-fit model to account for the X-ray background emission.
%We only used the MOS data because of their lower detector background level than the pn data.

\begin{deluxetable*}{cccc}[ht]
\tablecaption{Best-fit model parameters of the background spectrum.\label{tab:bkg_model}}
%\tablewidth{700pt}
%\tabletypesize{\scriptsize}
\tablehead{
\colhead{Physical Component} &
\colhead{XSPEC model} & 
\colhead{Parameter} &
\colhead{Value}
} 
\startdata
      GH & APEC (GH$_{\rm cold}$) & $kT_e$ (keV) & 0.658 (fixed) \\
      %& & $Z_{\rm all}$ (solar) &  1.0 (fixed) \\
      & & Norm\tablenotemark{a} &  $12.5\pm1.2$ \\
      & APEC (GH$_{\rm hot}$) & $kT_e$ (keV) & $1.22\pm0.02$ \\
      %& & $Z_{\rm all}$ &  1.0 (fixed)\\
      & & Norm\tablenotemark{a} &  $66.8^{+5.7}_{-6.9}$ \\  \hline
      LHB & APEC & $kT_e$ (keV) & 0.105 (fixed) \\
      %& & $Z_{\rm all}$ (solar) &  1.0 (fixed) \\
      & & Norm\tablenotemark{a} &  $45.1^{+7.7}_{-7.6}$ \\ \hline
      CXB & TBabs (Absorption) & ${N_{\rm H}}^{\rm CXB}$ & 0.60$^{+0.06}_{-0.05}$ \\
      & Power law & $\Gamma$ & 1.40 (fixed) \\
      & & Norm\tablenotemark{b} &  $12.1^{+1.3}_{-1.1}$  \\ \hline
       & & {$\chi^{2}_{\nu}$ ($\nu$)}\tablenotemark{c} & 1.44 (224) \\
\enddata
\tablenotetext{a}{The emission measure integrated over the line of sight, i.e., $(1/4\pi D^2) \int n_e n_{\rm H} dl$ in units of $10^{-14}$~cm$^{-5}$ ~sr$^{-1}$.}
\tablenotetext{b}{Units of photons s$^{-1}$ cm$^{-2}$ keV$^{-1}$ sr$^{-1}$ at 1~keV.}
\tablenotetext{c}{The parameters ${\chi_\nu}^2$ and $\nu$ indicate a reduced chi-squared and a degree of freedom, respectively.}
\label{tab:bkg_model}
%\begin{center}
%{
%$^{\ast}$The unit is photons s$^{-1}$ cm$^{-2}$ keV$^{-1}$ sr$^{-1}$ at 1~keV.\\
%$^{\dagger}$The emission measure integrated over the line of sight, i.e., (1 / 4$\pi D^2$) $\int n_e n_{\rm H} dl$ in units of 10$^{-14}$ cm$^{-5}$ sr$^{-1}$.\\ 
%$^{\S}$ The parameters ${\chi_\nu}^2$ and $\nu$ indicate a reduced chi square and a degree of freedom, respectively.
%}
%\end{center}
\end{deluxetable*}

\subsection{Spectral Analysis}\label{subsec:spec}

\begin{figure}[ht]
\begin{center}
 \includegraphics[width=7.0cm]{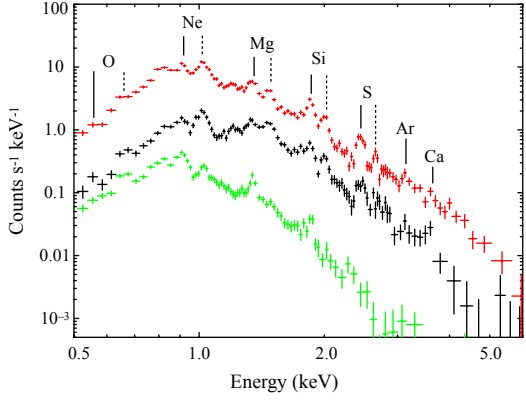} 
\end{center}
\vspace{-4mm}
\caption{
MOS (MOS1 + MOS2) spectra extracted from the representative subregions 1 (black), 2 (green), 3 (red) shown in Figure \ref{fig:Xray_image}(c). 
The NXB and X-ray background are subtracted. 
For a display purpose, the spectra of Regions 2 and 3 are scaled by factors of 0.2 and 4.0, respectively. 
The vertical solid and dashed lines denote the centroid energies of the He$\alpha$ lines and Ly$\alpha$ lines, respectively. 
}
\label{fig:all_spectrum}
\end{figure}

\begin{figure*}
\begin{tabular}{cccc}
\begin{minipage}[c]{0.5\hsize}
 \includegraphics[width=8cm]{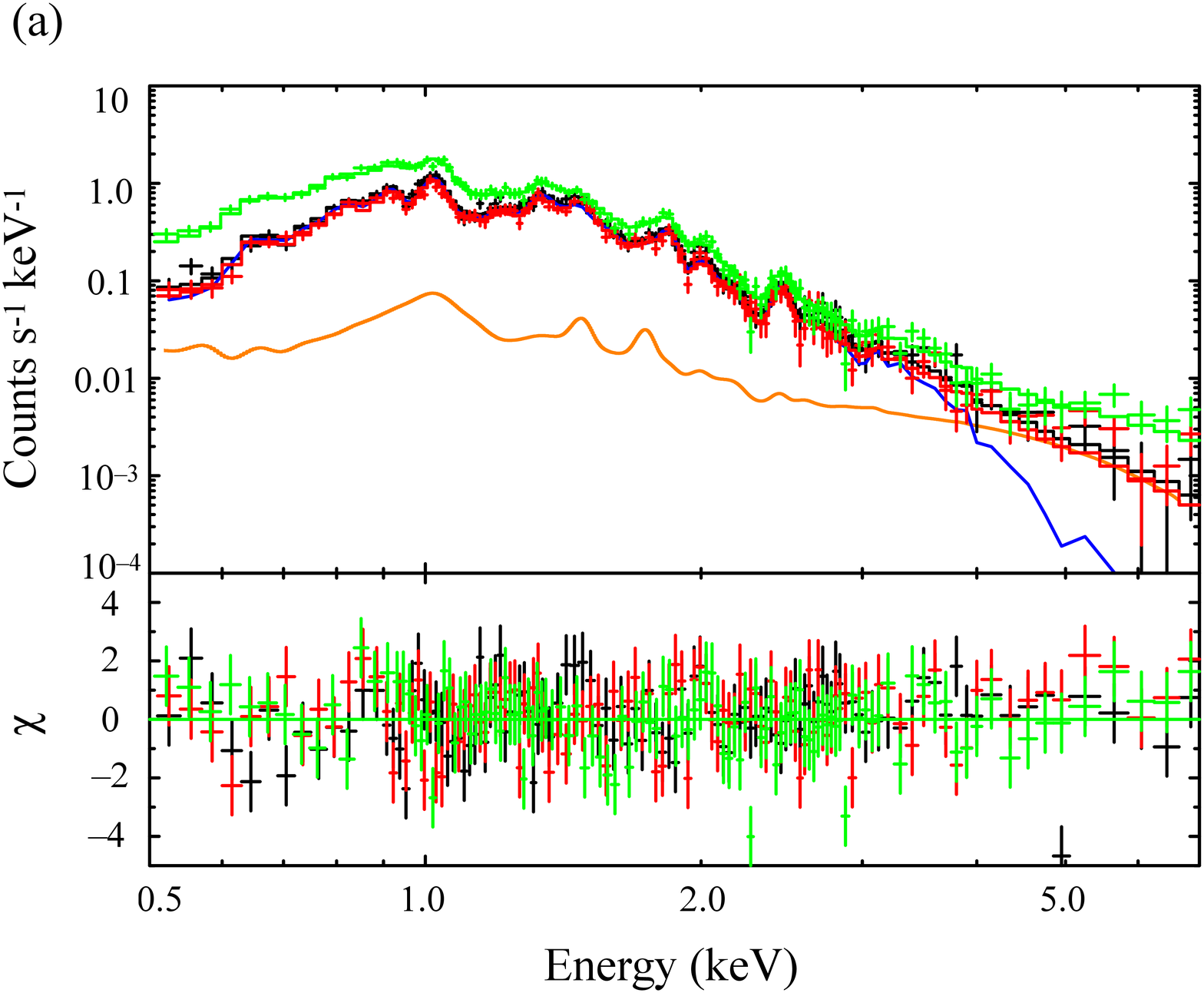} 
\end{minipage}
\begin{minipage}[c]{0.5\hsize}
 \includegraphics[width=8cm]{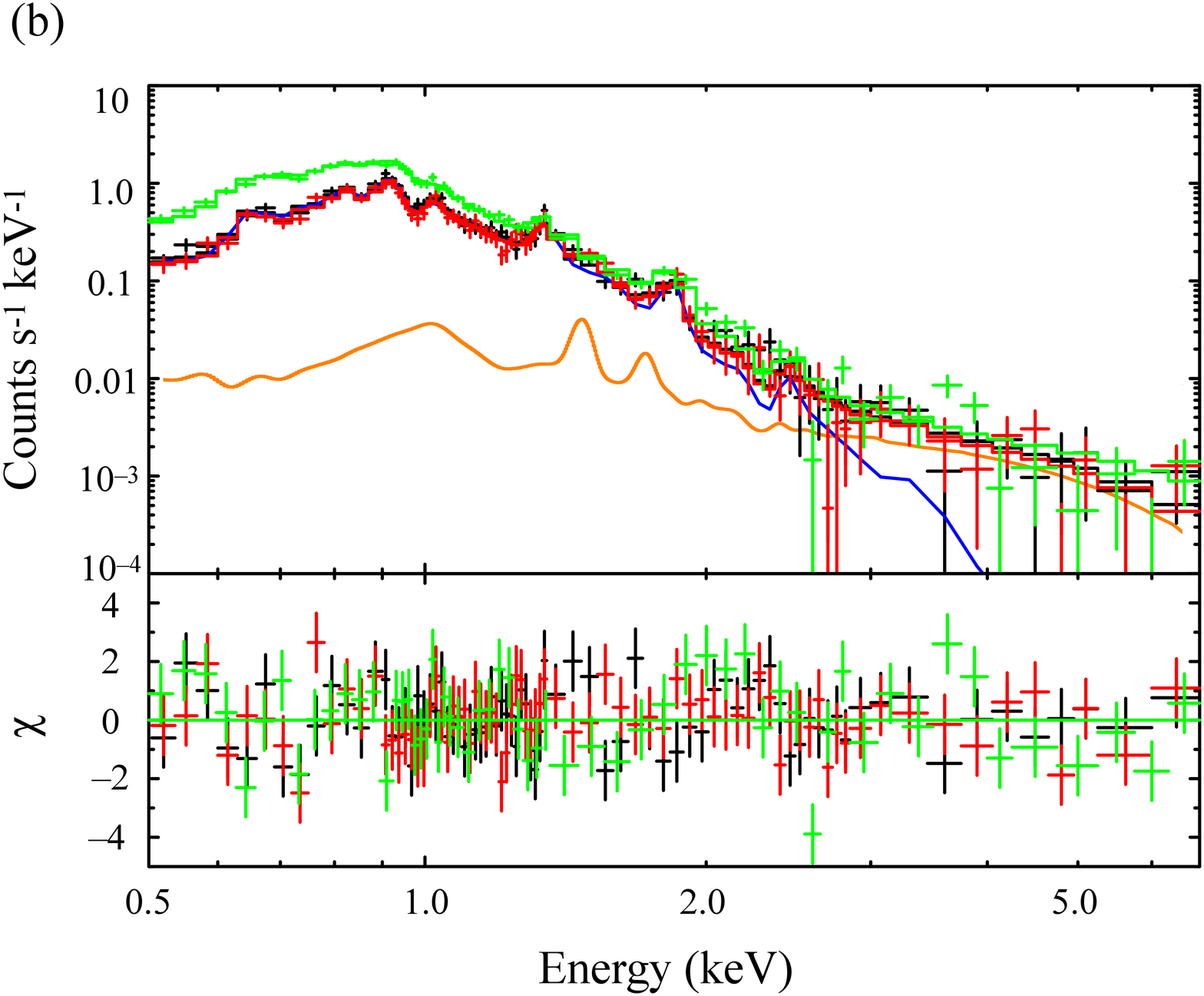} 
\end{minipage}
\\\\\\
\begin{minipage}[c]{0.5\hsize}
\vspace{-5mm}
 \includegraphics[width=8cm]{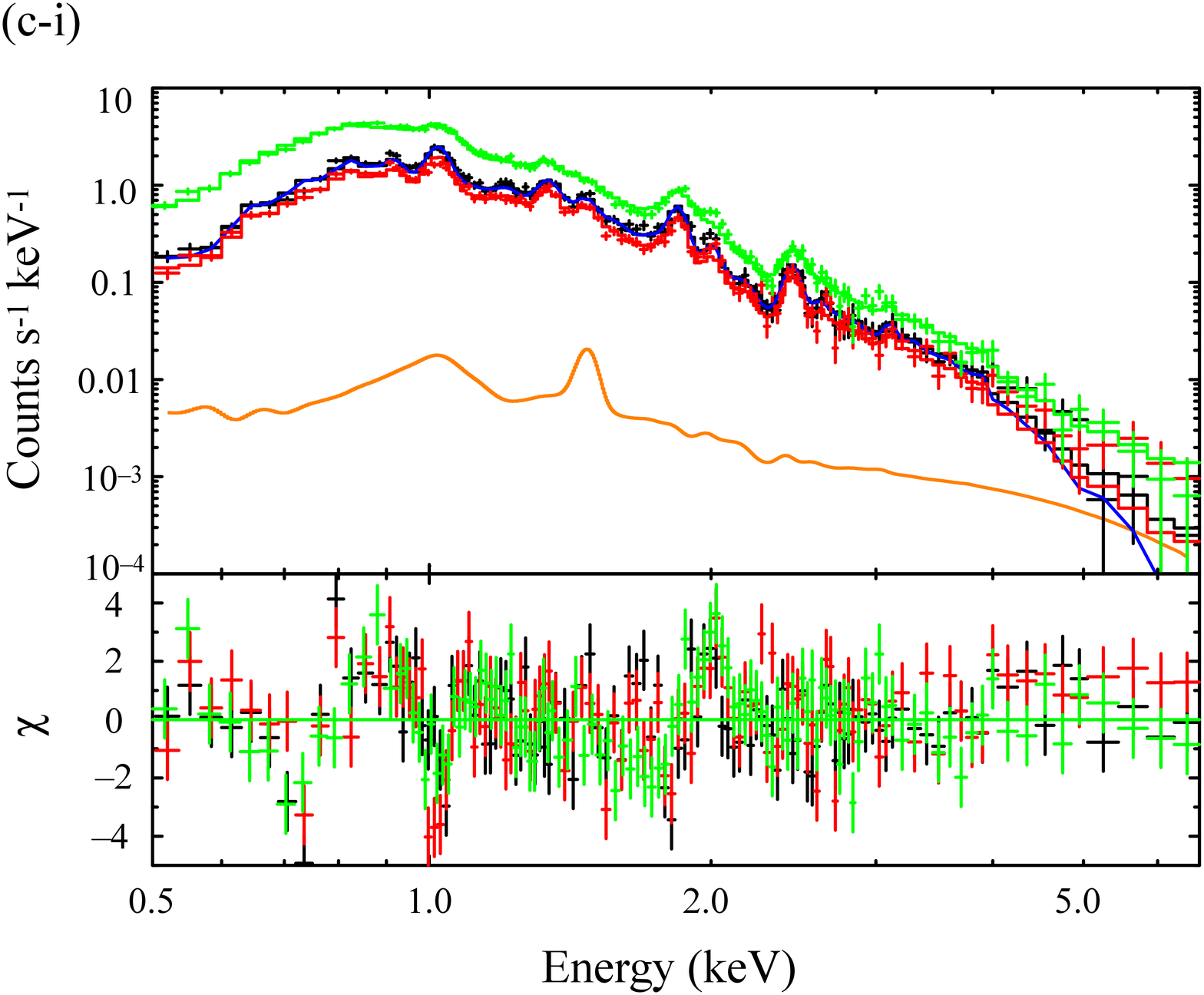} 
\end{minipage}
\begin{minipage}[c]{0.5\hsize}
\vspace{-5mm}
 \includegraphics[width=8cm]{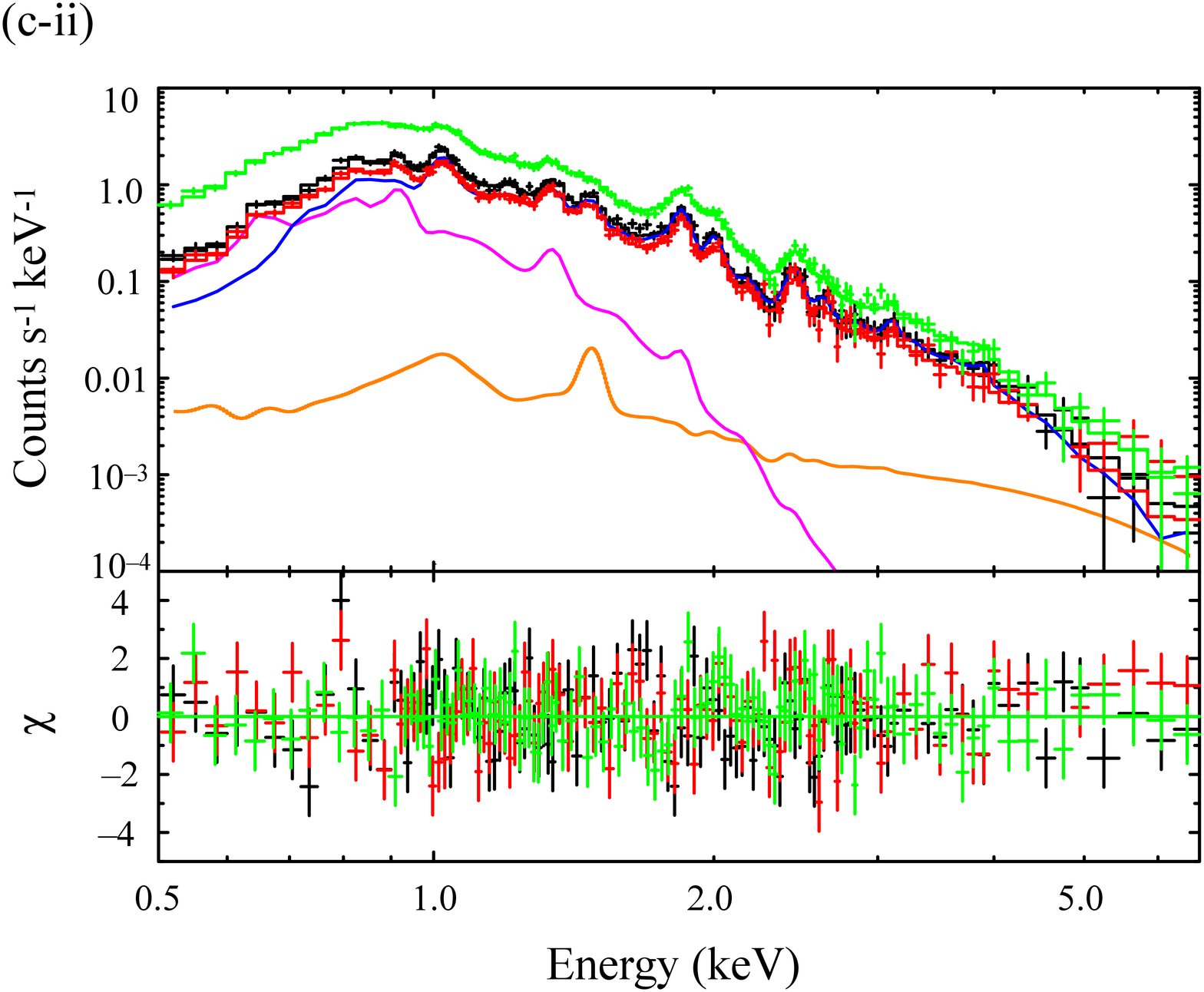} 
\end{minipage}
\end{tabular}
\vspace{0mm}
\caption{
(a) MOS1 (red), MOS2 (black), and pn (green) spectra from Region~1 plotted with the best-fit model. 
The blue, orange, and black curves represent the RP model, the background, and the sum of the models to MOS1 data, respectively. 
%(1-b) Same as panel (1-a) but with a Gaussian at 1.23 keV added to the model (the black dotted curve). 
%(1) Same as panel (1-a) but with a Gaussian at 1.23 keV added to the model (the black dotted curve). 
(b)--(c) Same as panel (a) but for Regions~2--3.
In the panel(c-ii), the magenta curve indicates the CIE component. 
%Since Region~3 is contaminated by the southern hard source (PSR B1853$+$01 and its PWN), the model includes a power law shown as the green curve. 
}
\label{fig:fittings}
\end{figure*}

\begin{figure}
\vspace{2mm}
\begin{center}
 \includegraphics[width=8cm]{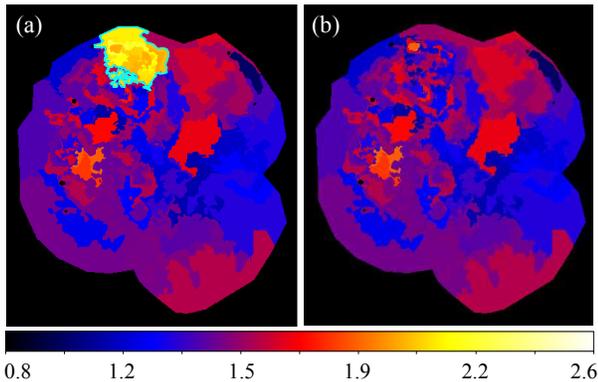} 
\end{center}
\vspace{-3mm}
\caption{
 (a) Reduced chi-squared map with the RP model.
  The cyan curve indicates the $\chi^2_\nu\geq2.0$ subregions.
 (b) Same as (a) but we applied the RP$+$CIE model to only the fits of the $\chi^2_\nu\geq2.0$ subregions   
}
\label{fig:chi}
\end{figure}

\begin{deluxetable*}{ccccccc}[ht]
\tablecaption{Best-fit model parameters of the spectra from the representative subregions.\label{tab:plasma_model}}
\tablehead{
%\colhead{Model function} &
\colhead{Physical Component} &
\colhead{XSPEC model} &
\colhead{Parameters} &
\colhead{Region 1} & 
\colhead{Region 2} &
\colhead{Region 3~(1RP)} &
\colhead{Region 3~(1RP+1CIE)}
} 
\startdata
       & TBabs & $N_{\rm H}$ (10$^{22}$ cm$^{-2}$) & 1.02$^{+0.3}_{-0.1}$ & 0.78$\pm$0.03 & 0.78$\pm$0.03 & 0.90$\pm$0.01 \\ 
       RP & VVRNEI & $kT_e$ (keV) & $0.24 \pm 0.02$ & 0.23$\pm$0.02& 0.36$\pm0.02$ & 0.53$\pm$0.02\\
       & & $kT_{\rm init}$ (keV) & 5.0 (fixed) & 5.0 (fixed) & 5.0 (fixed) & 5.0 (fixed) \\
       & & $Z_{\rm O}$ (Solar) & $0.5\pm{0.1}$ & 0.4$\pm$0.1 & 0.5$\pm0.1$ & < 1.5 \\
       & & $Z_{\rm Ne}$ (Solar) & 1.1$\pm0.2$ & 0.5$\pm$0.1 & 1.0$\pm$0.1 & 1.7$\pm$0.2 \\
       & & $Z_{\rm Mg}$ (Solar) & 0.8$\pm$0.1 & 0.5$\pm$0.2 & 0.8$\pm$0.1 & 1.2$\pm$0.1 \\
       & & $Z_{\rm Si}$ (Solar) & 0.5$\pm$0.1 & 1.0$\pm$-0.2 & 1.4$\pm$0.2 & 2.0$\pm$0.2 \\
       & & $Z_{\rm S}=Z_{\rm Ar}=Z_{\rm Ca}$ (Solar) & 1.2$^{+0.3}_{-0.2}$ & 1.0$\pm$0.1 & 1.1$\pm$0.2 & 1.3$\pm$0.2 \\ 
       & & $Z_{\rm Fe}=Z_{\rm Ni}$ (Solar) & 0.28$^{+0.05}_{-0.06}$ &0.3$\pm$0.1 & 0.27$^{+0.05}_{-0.04}$ & 0.5$\pm$0.1 \\
       & & $n_et$ (10$^{11}$ cm$^{-3}$s) & 4.6$\pm$0.2 & 12.7$^{+1.1}_{-1.2}$ & 6.0$\pm$0.2 & 5.5$^{+0.3}_{-0.2}$ \\
       & & Norm$^{a}$ & 0.30$\pm$0.01 &  0.14$\pm$0.02 & 0.14$\pm$0.02 & 0.07$\pm$0.01 \\  
       CIE & APEC & $kT_e$ (keV) & - & - & - & 0.21$^{+0.06}_{-0.07}$\\
       & & $Z_{\rm all}$ (Solar) & - & - & - & 1.0 (fixed) \\
       & & Norm$^{a}$ & - & - & - & 0.11$\pm$0.02 \\ \hline
      %Gaussian & Centroid (keV) & 1.23 (fixed) & 1.23 (fixed) & 1.23 (fixed) & 1.23 (fixed) \\
      %& Norm$^{\ast}$ & & & & \\ \hline
      & & {$\chi^{2}_{\nu}$ ($\nu$)}$^{b}$ & 1.19 (342) & 1.24 (201) & 2.04 (335) & 1.25 (333) \\
\enddata
%\tablenotetext{a}{The emission measure integrated over the line of sight, i.e., $(1/4\pi D^2) \int n_e n_{\rm H} dl$ in the unit of $10^{-14}$~cm$^{-5}$ ~sr$^{-1}$.}
%\tablenotetext{b}{The parameters ${\chi_\nu}^2$ and $\nu$ indicate a reduced chi-squared and a degree of freedom, respectively.}
%\tablenotetext{a}{RP component.}
%\tablenotetext{b}{ISM component.}
\tablenotetext{a}{The emission measure integrated over the line of sight, i.e., $(1/4\pi D^2) \int n_e n_{\rm H} dl$ in units of $10^{-14}$~cm$^{-5}$ ~sr$^{-1}$.}
\tablenotetext{b}{The parameters ${\chi_\nu}^2$ and $\nu$ indicate a reduced chi-squared and a degree of freedom, respectively.}
\end{deluxetable*}

\begin{deluxetable}{cccc}[ht]
\tablecaption{Average charge $\overline{C_{\rm RP}}$ and charge deviation $\Delta C$ for each ion species of RPs in the representative subregions.\label{tab:charge_fraction}}
\tablehead{
\colhead{Parameters} &
\colhead{Region 1} & 
\colhead{Region 2} &
\colhead{Region 3}
} 
\startdata
       $\overline{C_{\rm RP,\,O}}$ & 7.63$\pm$0.05 & 7.3$\pm$0.1 & 7.95$^{+0.03}_{-0.06}$ \\ 
       $\overline{C_{\rm RP,\,Ne}}$ & 9.22$\pm$0.04 & 8.4$\pm$0.1 & 9.66$\pm$0.01 \\ 
       $\overline{C_{\rm RP,\,Mg}}$ & 10.90$^{+0.04}_{-0.05}$ & 10.13$^{+0.06}_{-0.02}$ & 11.2$^{+0.02}_{-0.06}$ \\ 
       $\overline{C_{\rm RP,\,Si}}$ & 12.53$^{+0.05}_{-0.06}$ & 11.7$\pm$0.1 & 12.8$\pm$0.1 \\ 
       $\overline{C_{\rm RP,\,S}}$ & 13.8$\pm$0.2 & 12.4$^{+0.2}_{-0.4}$ &  14.5$^{+0.2}_{-0.3}$ \\ 
       $\overline{C_{\rm RP,\,Fe}}$ & 15.7$\pm$0.3 & 14.9$^{+0.1}_{-0.5}$ & 17.6$^{+0.1}_{-0.2}$  \\ \hline
       $\Delta C_{\rm O}$ & 0.27$^{+0.09}_{-0.7}$ & 0.09$^{+0.03}_{-0.04}$ & (0.7$^{+0.4}_{-0.1})$$\times$10$^{-5}$ \\ 
       $\Delta C_{\rm Ne}$ & 1.04$\pm$0.03 & 0.30$^{+0.02}_{-0.06}$ & 0.11$^{+0.02}_{-0.04}$ \\ 
       $\Delta C_{\rm Mg}$ & 0.93$^{+0.03}_{-0.05}$ & 0.16$^{+0.03}_{-0.04}$ & 0.70$^{+0.04}_{-0.05}$ \\ 
       $\Delta C_{\rm Si}$ & 0.72$^{+0.03}_{-0.05}$ & 0.06$^{+0.01}_{-0.02}$ & 0.62$^{+0.03}_{-0.04}$ \\ 
       $\Delta C_{\rm S}$ & 1.5$\pm$0.4 & 0.24$^{+0.05}_{-0.10}$ & 0.8$\pm$0.2 \\ 
       $\Delta C_{\rm Fe}$ & 0.4$\pm$0.1 & (2.8$^{+3.3}_{-0.8}$)$\times$10$^{-5}$ & 1.0$^{+0.1}_{-0.2}$ \\ 
\enddata
\end{deluxetable}

Previous X-ray studies \citep[e.g.,][]{Matsumura2017b,Greco2018} revealed the presence of RPs almost in the whole of IC~443.
%Previous X-ray studies \citep[e.g.,][]{Matsumura2017b,Greco2018} revealed that RPs almost over the entire the remnant.
Figure~\ref{fig:all_spectrum} shows the spectra extracted from the representative subregions in Figure~\ref{fig:Xray_image}(c).
One can clearly see resolved emission lines from highly ionized O, Ne, Mg, Si, S, and Ar ions, as well as their different Ly$\alpha$/He$\alpha$ ratios and distinct continuum shape between each spectrum.
The spectral features indicate significant spatial variations of parameters of the RPs and of absorption column densities toward the remnant. 

To model the plasma emission, we used the VVRNEI \citep{Foster2017} model implemented in the XSPEC software version 12.10.1f \citep{Arnaud1996}.
The VVRNEI model can calculate the spectrum of thermal plasma in the overionized state after an abrupt decrease in the electron temperature.
The described ionization state is time-evolved with the recombination ``fluence'' $n_et$ after the cooling from $kT_{\rm init}$ to $kT_e$ under an assumption that the plasma initially was in a CIE state.
To take account of photoelectric absorption by the foreground gas, we applied the TBabs model.
We allowed the column density $N_{\rm H}$, the present electron temperature $kT_e$, density-weighted recombination time, $n_et$, and normalization of the VVRNEI component to vary.
The initial plasma temperature $kT_{\rm init}$ is constrained to be $\gtrsim$ 5~keV, whose elements such as O--S are almost fully ionized.
%The initial plasma temperature $kT_{\rm init}$ is constrained to be $\gtrsim$ 5~keV, in which ions such as O--S, except for Fe, are almost fully ionized.
We fixed $kT_{\rm init}$ at 5 keV because other parameters such as $N_{\rm H}$, $kT_e$, and $n_et$ are hardly sensitive to the choice of $kT_{\rm init}$. 
We let the abundances of O, Ne, Mg, Si, S, Ar, Ca, Fe and Ni vary, and tied Ar and Ca to S, and Ni to Fe.
%The abundances of O, Ne, Mg, Si, S, and Fe were left free, whereas Ar and Ca were linked to S, and Ni was linked to Fe. 
The abundances of the other elements were fixed to solar values.
In spectral fittings of the subregions where the PWN 1SAX J0617.1$+$2221 or its pulsar are observed (Figure~\ref{fig:Xray_image}(c)), we applied a model that consists of the VVRNEI model and an additional power law to account for their emissions.
The photon index $\Gamma$ and the normalization of the power law were left free.

Figures~\ref{fig:fittings} show the results of the spectral fittings of Regions 1, 2, and 3.
The VVRNEI model describing an RP (with the additional power law) gives good fits to spectra from most of subregions including Regions 1 and 2 (Figures~\ref{fig:fittings}(a) and 4(b)).
On the other hand, as represented by the example of Region 3 (Figure~\ref{fig:fittings}(c-i)), some fits left remarkable residuals at $\sim$~1.0 keV around \ion{Ne}{11} Ly$\alpha$ line and $\sim$~2.0~keV around \ion{Si}{15} Ly$\alpha$ line.
These residuals suggest plasma parameters are different between the soft and hard bands.
We followed the same modeling procedure as \cite{Matsumura2017b}, who reported the presence of a cooler ISM in addition to the hot RP components, and refitted the spectra from subregions with $\chi^2_\nu\geq2.0$ as indicated by the $\chi^2_\nu$ map in Figure~\ref{fig:chi}(i).
The spectra are well reproduced by the RP model with a CIE component (APEC) for the ISM origin.
In Figure~\ref{fig:fittings}(c-ii), we show the result of Region 3.

We also tried the NEI and PSHOCK models that are possible candidate models for the ISM component, instead of the APEC model to examine other model combination well representing the overall spectra.
The combination of the RP and either model successfully reproduces the spectra but does not significantly improve the fitting statistics compared to that including the APEC model (For Region 3, VVRNEI+NEI: ${\chi^2_\nu}$ = 1.25 with $\nu$ = 332; VVRNEI+VSHOCK: ${\chi^2_\nu}$ = 1.24 with $\nu$ = 332; VVRNEI+APEC: ${\chi^2_\nu}$ = 1.25 with $\nu$ = 333).
The choice of the different ISM model does not change the RP parameters obtained beyond the 90\% confidence level so that we used the model that consists of the VVRNEI and APEC.
All subregions finally have $\chi^2_\nu\leq2.0$ as shown in the $\chi^2_\nu$ map (Figure~\ref{fig:chi}(ii)).
The best-fit parameters of Regions 1--3 are summarized in Table \ref{tab:plasma_model}. 
%The charge fraction for each ion species which are parameters which enable us to directly evaluate the overionization degree in RPs.

To directly quantify the overionization degree in RPs, we introduce the average charge of each ion species, $\overline{C}$, and the deviation of the average charge from the CIE state with the same $kT_e$, $\Delta C$,
\begin{equation}
\overline{C} = \sum c_i\,F_i
\end{equation}
\begin{equation}
\Delta C = \overline{C_{\rm RP}}(kT_e, n_et, kT_{\rm init} = 5~{\rm keV}) - \overline{C_{\rm CIE}}(kT_e),
\end{equation}
where $c_i$ and $F_i$ are the charge number and fraction of $i$-times ion, respectively. 
We computed $F_i$ by using PyAtomDB\footnote{https://atomdb.readthedocs.io}.
$\overline{C}$ and $\Delta C$ of Regions 1--3 are listed in Table \ref{tab:charge_fraction}.

\section{Discussion}

\subsection{Foreground Gas Distribution}

\begin{figure}
\begin{center}
 \includegraphics[width=6cm]{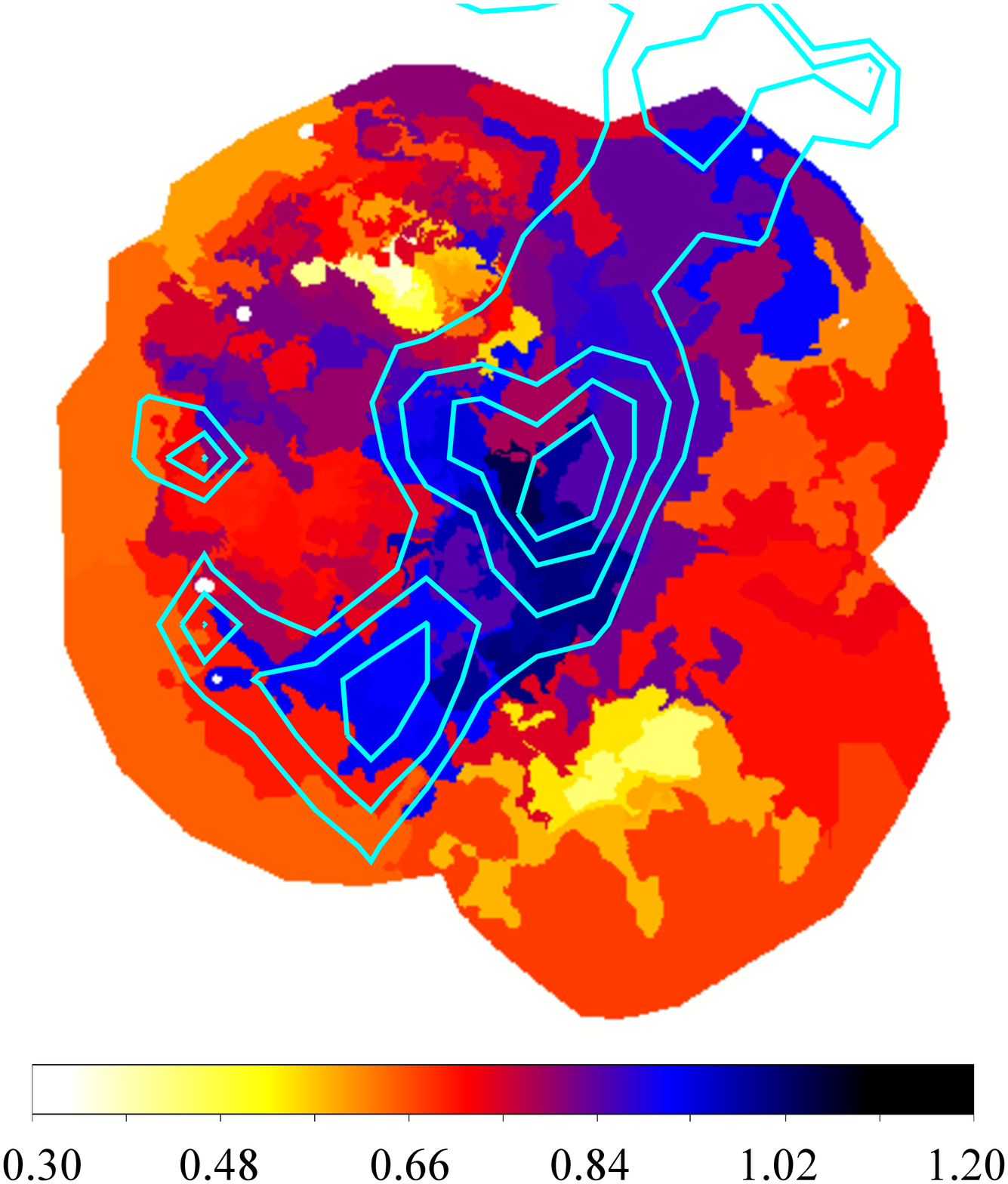} 
\end{center}
\vspace{-4mm}
\caption{
Distribution of X-ray absorption column density ($N_{\rm H}$). 
%The same radio continuum image as that in Figure~\ref{fig:Xray_image}(a) are overlaid as cyan contours in panel (a). 
The cyan contours denote $^{12}$CO($J=1$--$0$) emissions in a velocity range of $V_{\rm LSR} = -18$--$+2~{\rm km}~{\rm s}^{-1}$ as observed with the NANTEN2 \citep{Yoshiike2017}. 
}
\label{fig:nH}
\end{figure}

The foreground gas absorption $N_{\rm H}$ serves to probe the spatial distribution of the gas in front of the remnant.
Figure~\ref{fig:nH} shows the $N_{\rm H}$ values of each subregion.
We overlay the $^{12}{\rm CO}(J=1-0)$ emission observed with the NANTEN2 \citep{Yoshiike2017}.
The $N_{\rm H}$ map reveals higher values in subregions where the $^{12}{\rm CO}$ emission is detected.
%The morphological match was pointed out first by \cite{Matsumura2017b} and supports an interpretation that most of the gas traced by the CO line present in front of IC 443.
The spatial match was pointed out first by \cite{Matsumura2017b} and supports an interpretation that most of the gas traced by the CO line is present in front of IC 443.
Does the amount of gas expected by the spatial variation of $N_{\rm H}$ account for the CO data?
We roughly estimated the column density of the foreground gas to be $\sim0.6\times10^{22}~{\rm cm^{-2}}$ with the difference of the $N_{\rm H}$ values of the $^{12}$CO line observed and not observed subregions.
\cite{Yoshiike2017} estimate the column density $N_{\rm H,~CO}\sim0.6\times10^{22}~{\rm cm^{-2}}$ with the NANTEN2 data. 
Given that the amount of the atomic gas traced by HI emissions is less than that of the gas traced by the CO line \citep{Yoshiike2017}, our measurement is consistent with the radio one.

\subsection{Physical Origin of RPs}

\begin{figure*}
\begin{center}
 \includegraphics[width=15cm]{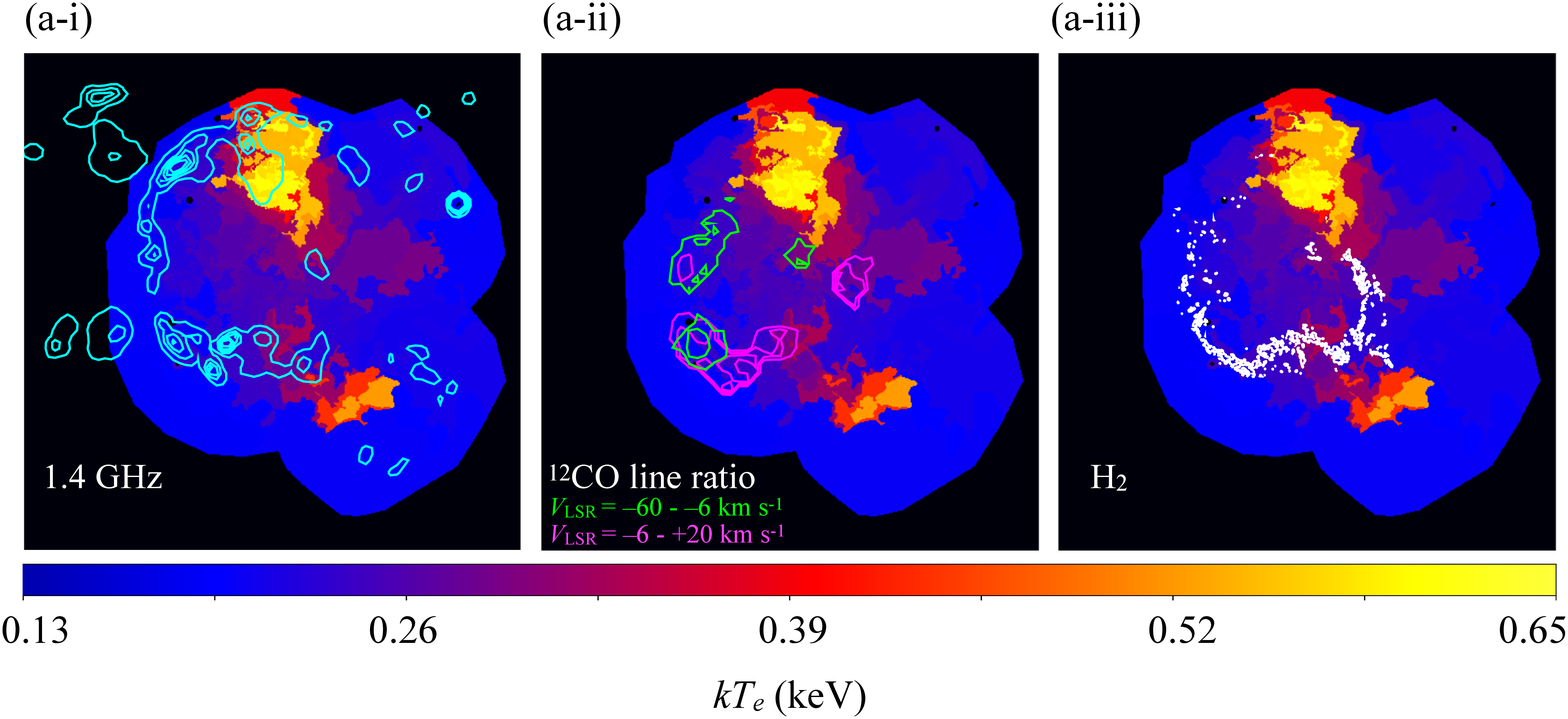} 
\end{center}
\begin{center}
  \includegraphics[width=15cm]{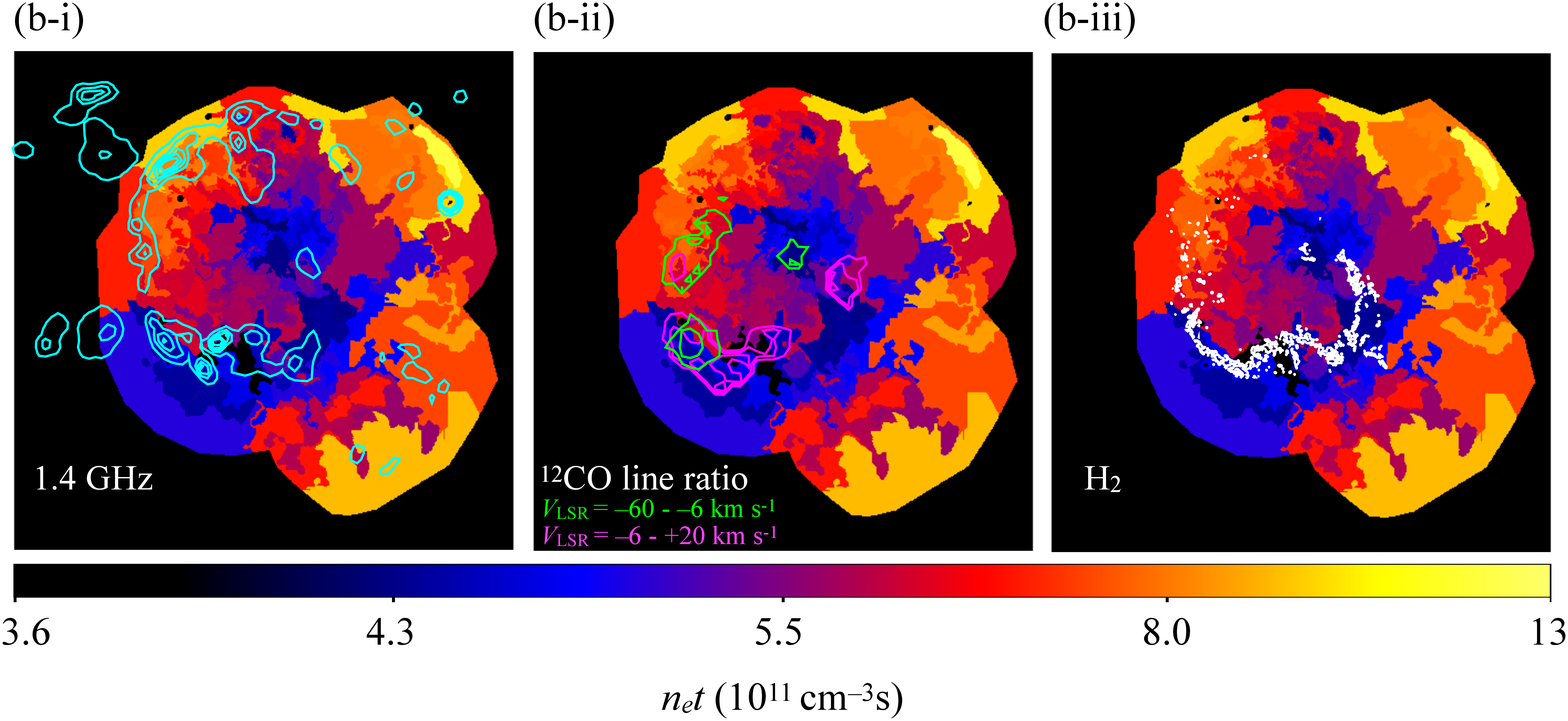} 
\end{center}
\vspace{-4mm}
\caption{
Maps presenting the distributions of (a) $kT_e$ and (b) $n_et$. 
The cyan contours in the panels (a-i) and (b-i) indicate the same radio continuum image as that in Figure~\ref{fig:Xray_image}(a). 
The green and magenta contours in panels (a-ii) and (b-ii) denote a $^{12}$CO($J=2$--$1$)-to-$^{12}$CO($J=1$--$0$) intensity ratio map drawn every 0.8 from 1.0 in $V_{\rm LSR} = -6$--$+20~{\rm km}~{\rm s}^{-1}$ and  $V_{\rm LSR} = -60$--$-6~{\rm km}~{\rm s}^{-1}$ taken from \cite{Yoshiike2017}. 
In panels (a-iii) and (b-iii), the white contours denote ${\rm H}_2$ 1--0 S(1) line with the InfraRed Survey Facility 1.4 m telescope by \cite{Kokusho2020}.
}
\label{fig:kTe_net_distribution}
\end{figure*}

\begin{figure}
\begin{center}
 \includegraphics[width=7.5cm]{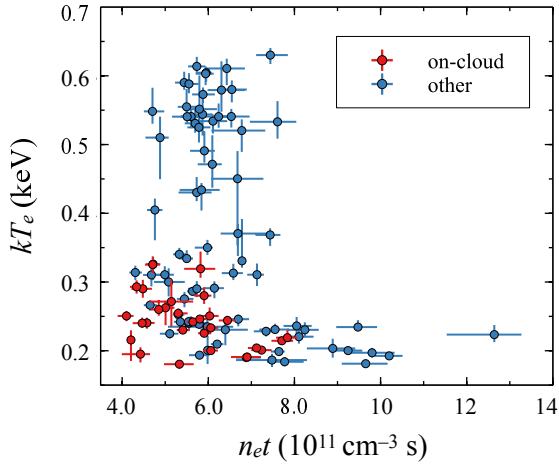} 
\end{center}
\vspace{-4mm}
\caption{
Relationship between $n_et$ and $kT_e$.
The red points are from on-cloud subregions where the $^{12}$CO line ratio or H$_2$ emission contours are superposed in maps in Figure~\ref{fig:kTe_net_distribution}, whereas the blue points are from other subregions. 
}
\label{fig:kTe-net}
\end{figure}

\begin{figure*}
\begin{center}
 \includegraphics[width=16cm]{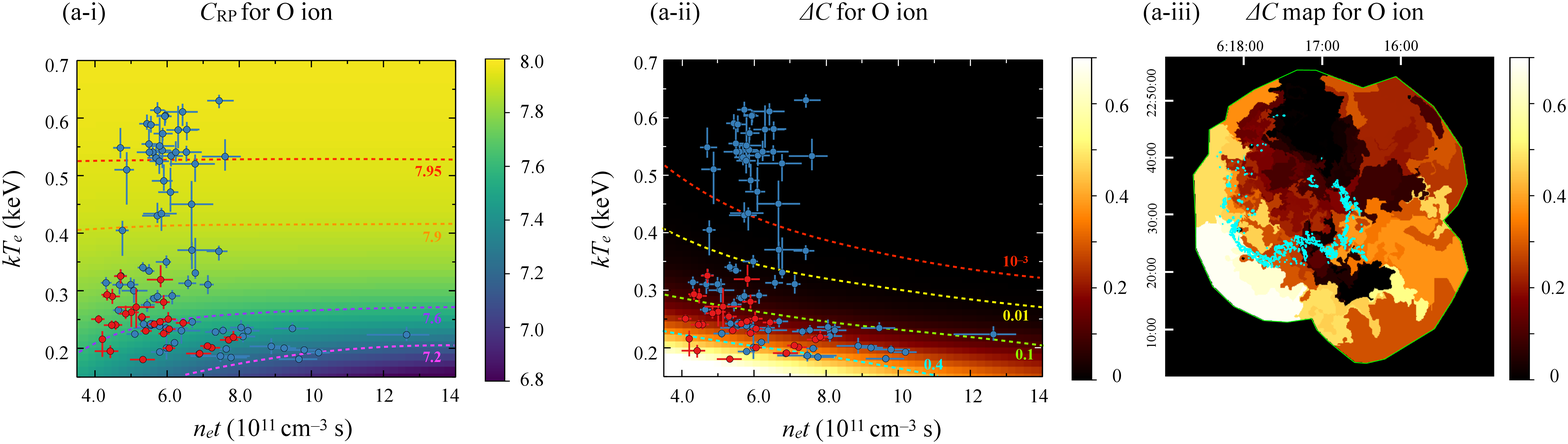} 
 \vspace{+2mm}
 \includegraphics[width=16cm]{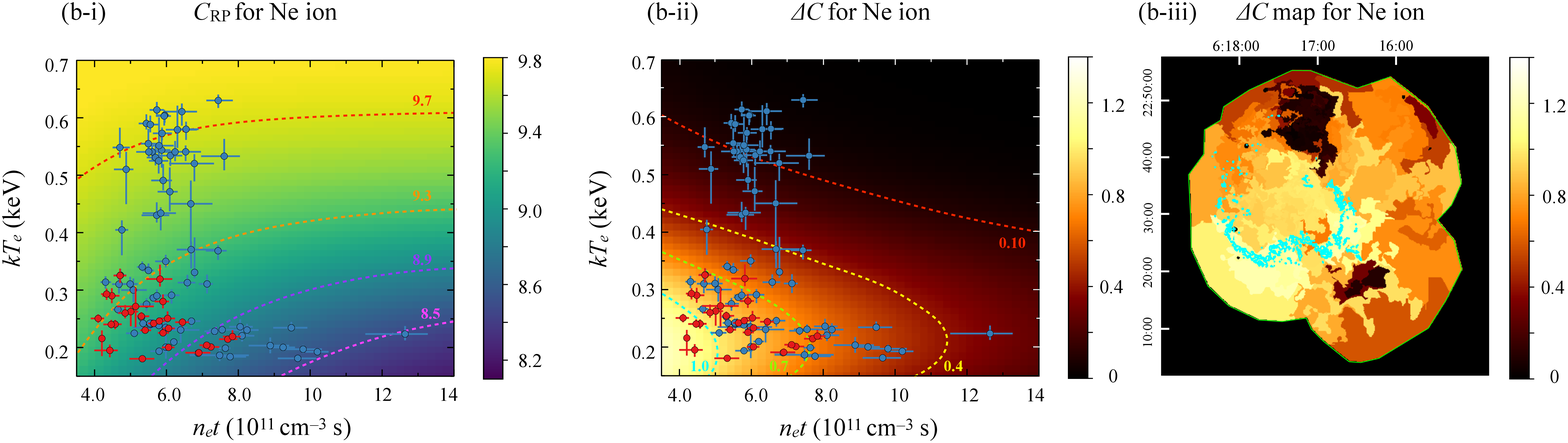} 
  \vspace{+2mm}
 \includegraphics[width=16cm]{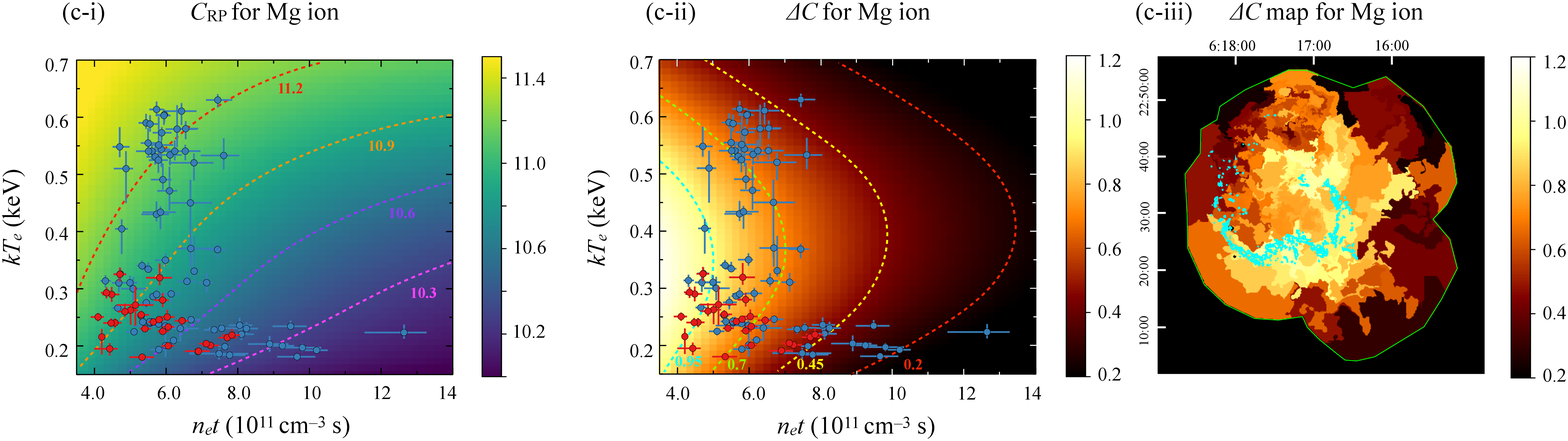} 
  \vspace{+2mm}
\end{center}
\vspace{-4mm}
\caption{
Average charge $C_{\rm RP}$ and charge deviation $\Delta C$ for O--Mg ions in the panels (a-i)- (c-i) and (a-ii)-(c-ii), respectively, shown on color scales.
We overlay the same plot as that in Figure~\ref{fig:kTe-net} on each map.
In panels (a-iii)-(c-iii), maps of $\Delta C$ computed with best-fit values.
The cyan contours denote the same radio ${\rm H}_2$ image as that in Figure~\ref{fig:kTe_net_distribution}.
}
\label{fig:average_charge_part1}
\end{figure*}

\begin{figure*}
%\addtocounter{figure}{-1}
\begin{center}
 \includegraphics[width=16cm]{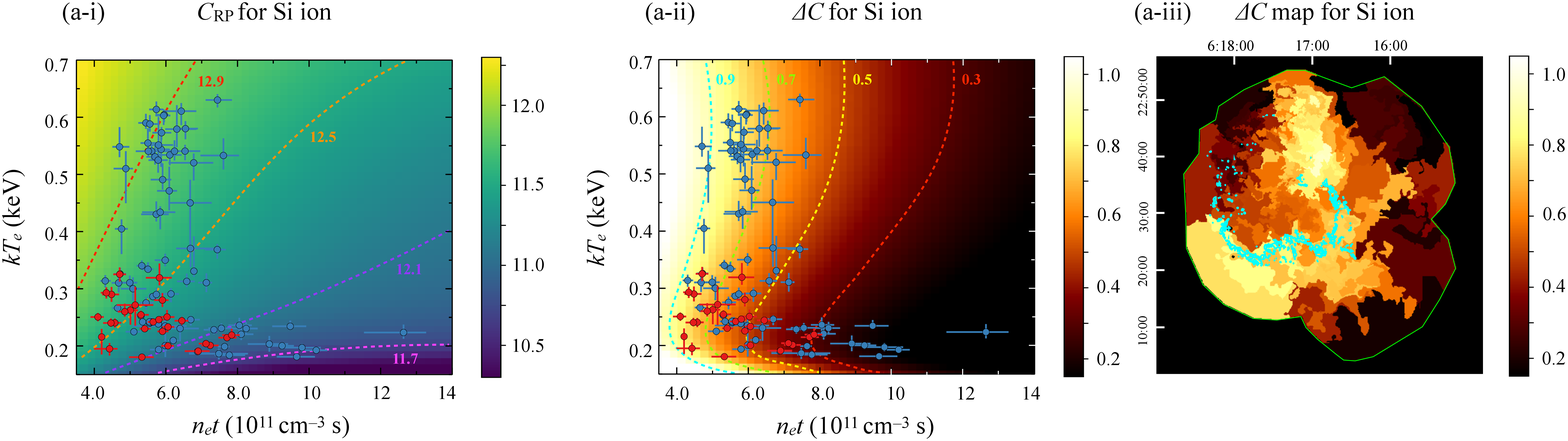} 
  \vspace{+4mm}
 \includegraphics[width=16cm]{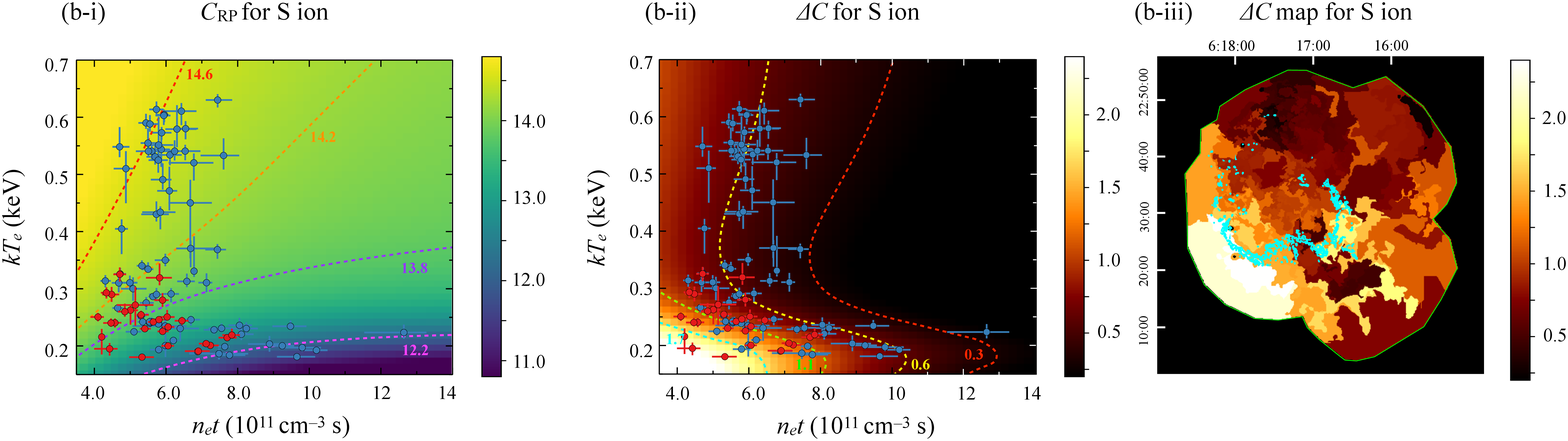} 
  \vspace{+4mm}
  \includegraphics[width=16cm]{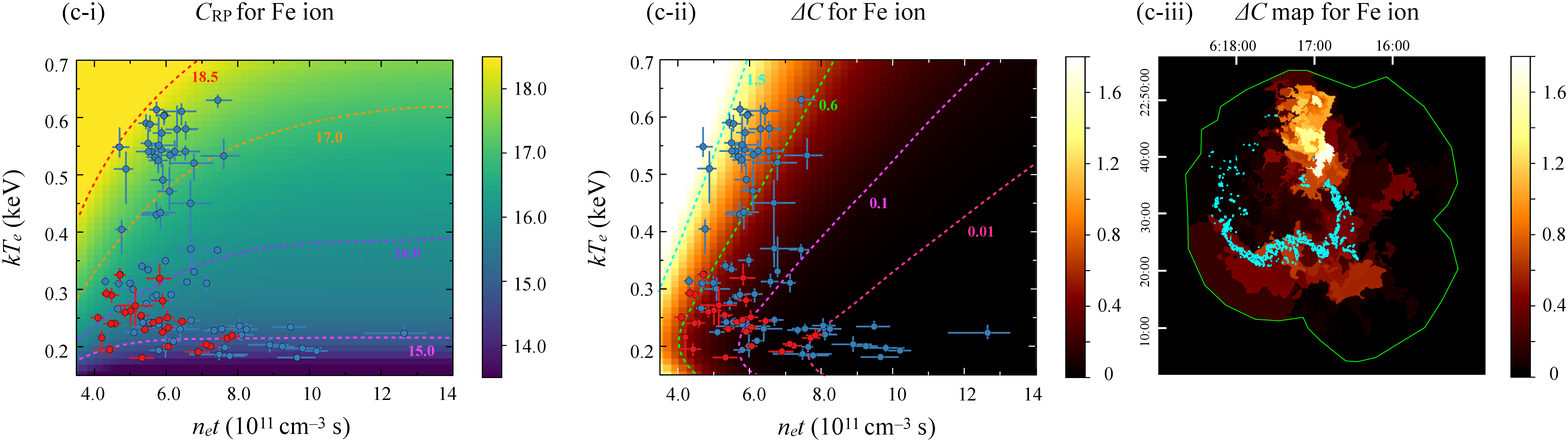} 
\end{center}
\vspace{-4mm}
\caption{
Same as Figure~\ref{fig:average_charge_part1} but for Si--Fe ions.
}
\label{fig:average_charge_part2}
\end{figure*}

We first focus on the thermal conduction and rarefaction scenarios, which are potentially responsible for the overionization in IC~443.
Spatial comparisons between $kT_e$, $n_et$, and the shocked clouds give crucial information for disentangling the two scenarios \citep{Yamaguchi2018,Okon2020}.
We map the $kT_e$ and $n_et$ derived from subregions (Figure~\ref{fig:kTe_net_distribution}) and plot them (Figure~\ref{fig:kTe-net}).
The $^{12}{\rm CO}$($J$ = 2--1) -to- $^{12}{\rm CO}$($J$ = 1--0) line ratio and ${\rm H}_2$1--0 $\it S($1) line contours are superposed on the maps to clarify the locations of shocked clouds.
Based on these maps, we color data points from on-cloud subregions, which overlap with either of the $^{12}$CO or ${\rm H}_2$ emission, in red in the plot.
We find decreased $kT_e$ and $n_et$ of RPs toward the on-cloud subregions. 

The physical implication of the gradients of $kT_e$ and $n_et$ in RPs can be understood by investigation of the charge fraction for each ion.
In Figures~\ref{fig:average_charge_part1} and \ref{fig:average_charge_part2}, the average charge $\overline{C_{\rm RP}}$ and the charge deviation $\Delta C$ are shown on the same $kT_e$-$n_et$ plot as Figure~\ref{fig:kTe-net}. 
%We also presented the maps of $\Delta C$ obtained in the same manner of that of $kT_e$ and $n_et$.
We also present the maps of $\Delta C$.
%The maps of $\Delta C$ are also obtained in the same manner of that of $kT_e$ and $n_et$.
RPs with lower $kT_e$ and $n_et$ have larger $\Delta C$ particularly for Ne--S ions. 
Our new result indicates RPs in the on-cloud subregions tend to be more cooled and more strongly overionized.
In the context of the thermal conduction scenario claimed by \cite{Matsumura2017b}, these tendencies are naturally interpreted as the result of a rapid cooling in the region where the shock is interacting with clouds.
On the other hand, adiabatic expansion in the rarefaction scenario favored by \cite{Greco2018}, requires higher $kT_e$ and smaller $\Delta C$ (or larger $n_et$) in the on-cloud subregions with a high density, which is inconsistent with our result.
In order to discuss the overionization degree of RPs in more detail, spectral analysis taking account of realistic SNR evolution would be helpful.
For instance, we assumed that the plasma initially was in a CIE state and its ionization state is uniform in the whole remnant, but the two assumptions would not hold in reality \cite[e.g.,][]{Sawada2012}.
 
It is interesting to point out that our result for IC~443 is similar to that on W44, whose RP is also ascribed to thermal conduction \citep{Okon2020}.
Regions with lower $kT_e$ and $n_et$ in W44 completely coincide with the locations where a spatially extended broad $^{12}$CO line, the so-called ``SEMBE'', is observed.
Although the nature of SEMBE is not clear yet, the emission is considered to be from small clumps \cite[$\ll$ 0.3 pc;][]{Sashida2013} disturbed after the shock propagation \citep{Seta2004,Sashida2013}.
\cite{Okon2020} claimed that hot plasma is efficiently cooled by evaporation of the clumps embedded in the plasma via thermal conduction.
The same mechanism may account for the $kT_e$ and $n_et$ trends in IC~443.
A deep and high angular resolution CO mapping would help to search for the spatially-extended broad line structures.
Comparison between the overionization degree of RPs and  the properties of the shocked gas can provide a step forward in the study of the cooling mechanism via thermal conduction. 
%Comparison between the properties of RPs revealed by X-ray observations and the radio observations can make a step forward in the study of the cooling mechanism via thermal conduction.

\begin{figure}
\begin{center}
 \includegraphics[width=6.5cm]{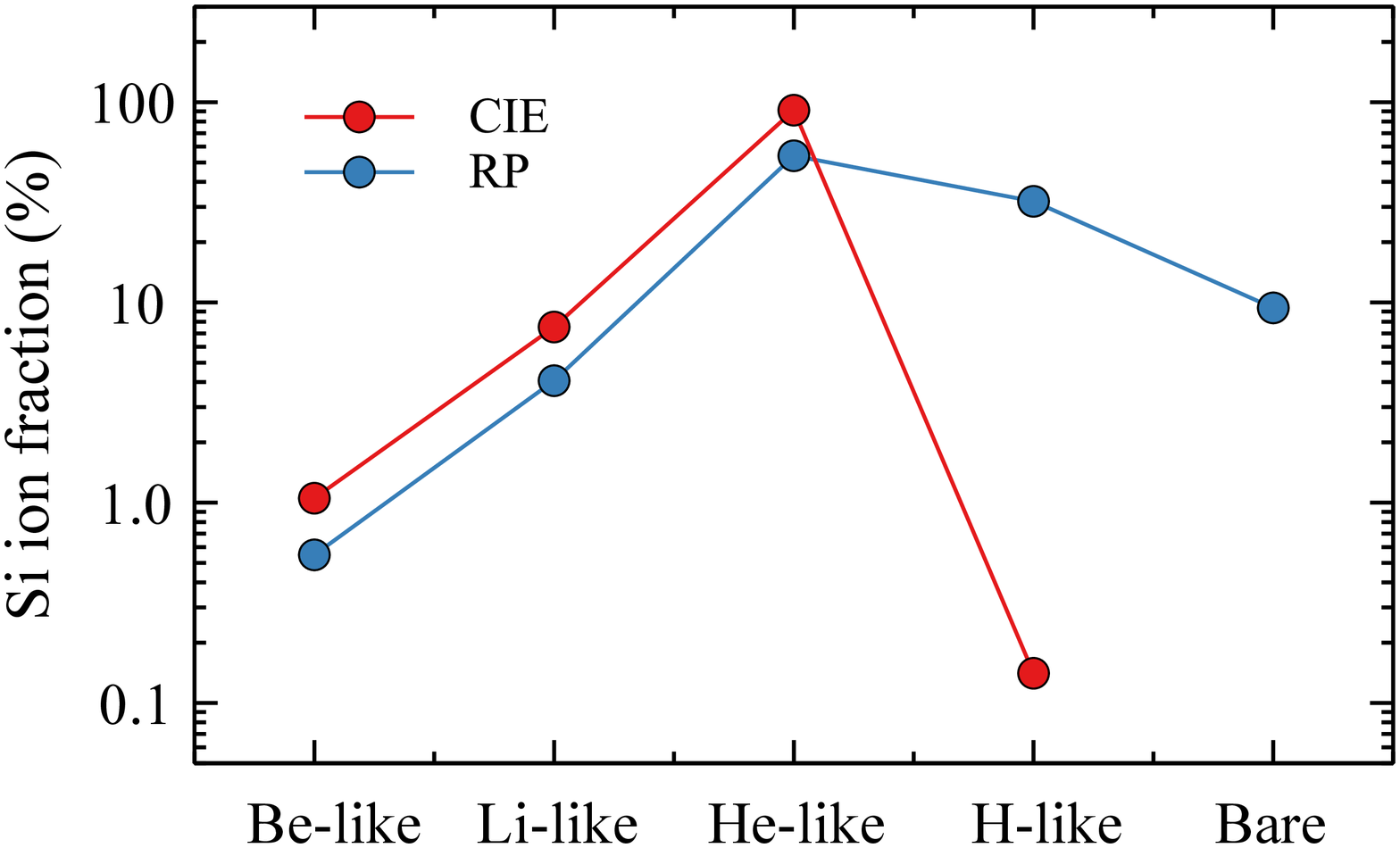} 
\end{center}
\vspace{-4mm}
\caption{
Si ion fraction in a CIE plasma (red) and an RP (blue).
RP parameters are typical obtained values of RPs in IC 443, $kT_{\rm init} = 5.0~{\rm keV}$, $kT_e= 0.3~{\rm keV}$, and $n_et = 6.0\times10^{11}~{\rm cm^{-3}s}$.
The CIE plasma has the same $kT_e= 0.3~{\rm keV}$ as that of the RP model.
}
\label{fig:ion_pop}
\end{figure}

%\subsection{Association between RPs and protons accelerated in SNRs}
\cite{Hirayama2019} and \cite{Yamauchi2021} proposed a new scenario where the overionization in SNRs is caused by bombardment of accelerated protons.
Given ionization cross sections, sub-relativistic protons may more efficiently ionize ions than thermal electrons in the plasma.
We test the possibility from the viewpoint of the energetics of particle acceleration.
Figure~\ref{fig:ion_pop} shows the Si ion population in the RP with typical parameters obtained, $kT_e=0.3~{\rm keV}$, $n_et=6.0\times10^{11}~{\rm cm^{-3}s}$, and $kT_{\rm init} = 5.0~{\rm keV}$, and that in a CIE plasma with the same $kT_e$.
To explain the strong H-like Si RRC in the IC~443 spectra, which is solid evidence for the RPs \citep[e.g.,][]{Yamaguchi2009}, some of the abundant He-like Si ions in the CIE plasma must be ionized to the H-like and subsequently to fully ionized states.
%abundant He-like Si ion in the CIE plasma must be ionized to H-like Si ion as well as subsequently to fuli ionized.

We now estimate the required energy density of sub-relativistic protons under the condition that the ionization of He-like to H-like Si ions by the proton bombardment proceeds faster than the relaxation toward a CIE state. 
The ionization rate $\xi$ can be described as
%The ionization rate $\xi$ and the energy density $\epsilon_{\rm p}$ can be described as follows,
\begin{equation}
\xi = \int \sigma_{\rm i}\,v\,\frac{dn_{\rm p}}{dE}dE, \label{eq:rate}
\end{equation}
where $\sigma_{\rm i}$ is the K-shell ionization cross section for He-like Si ions.
Assuming the proton spectrum $dn_{\rm p}/dE\propto E^{-2}$ expected in diffusive shock acceleration at a strong shock and $\sigma_{\rm i}$ by \cite{McGuire1973}, $\xi$ is estimated as
\begin{equation}
\xi \sim 3.3\times\left(\frac{n_{\rm p}}{1.0~{\rm cm^{-3}}}\right)\times10^{-12}~{\rm s^{-1}}, \label{eq:rate2}
\end{equation}
where $n_p$ is the proton density integrated in the energy range from $0.2~{\rm MeV}$ to $20~{\rm MeV}$ corresponding to the integral ranges of Equation \ref{eq:rate}.
\cite{Smith2010} gave the characteristic timescale toward CIE as
\begin{equation}
\Lambda\sim 3\times\left(\frac{n_e}{1.0~{\rm cm^{-3}}}\right)^{-1}\times10^{11}~{\rm s}. \label{eq:relaxation}
\end{equation}
To meet the condition, $\xi$ must be larger than $\Lambda^{-1}$.
If $n_e$ is $1.0~{\rm cm^{-3}}$, $n_{\rm p}$ is estimated to be $\geq 1.0~{\rm cm^{-3}}$.
The energy density $\epsilon_{\rm p}$ and total energy $E_{\rm p}$ can be described as
\begin{equation}
\epsilon_{\rm p} = \int E\,\frac{dn_{\rm p}}{dE}dE \sim 0.9\times\left(\frac{n_{\rm p}}{1.0~{\rm cm^{-3}}}\right)~{\rm MeV\,cm^{-3}}, \label{eq:density2}
\end{equation}
\begin{equation}
E_{\rm p} =\epsilon_{\rm p}\cdot V \cdot f,
\end{equation}
where $V$ and $f$ are the volume of the whole IC~443 and the filling factor, respectively. 
Assuming the volume is a sphere with a radius of $\sim$10~pc, $E_{\rm p}$ is estimated to be $\sim1.8\cdot f\cdot ({n_{\rm p}}/{1.0~{\rm cm^{-3}}})\cdot 10^{53}$ erg.
The proton energy certainly exceeds the typical kinetic energy ($\sim$10$^{51}$ erg) in supernova explosions even if we consider the uncertainity of $f$.
Our estimation thus indicates that it is difficult to explain the observed RPs in IC~443 only by the proton ionization claimed by \cite{Hirayama2019} and \cite{Yamauchi2021} and the contribution is negligible.

\section{Conclusions}

We have performed spatially resolved spectroscopy of the X-ray emission of IC~443 with {\it XMM-Newton}, aiming to clarify the physical origin of the overionization.
All spectra extracted from each region are well fitted with an RP model or the RP model with an additional CIE model of shocked ISM origin.
The X-ray absorption column density is higher in the region with the bright $^{12}$CO line emission, indicating that the gas traced by the CO line is present in front of IC~443.
The obtained electron temperature $kT_e$ and the recombining degree $n_et$ of RPs range from 0.15~keV to 0.65~keV and from $4.0\times10^{11}~{\rm cm^{-3}\,s}$ to $14\times10^{11}~{\rm cm^{-3}\,s}$, respectively.
We have discovered that RPs in the region where the shock is interacting with ambient clouds tend to have lower $kT_e$ and smaller $n_et$.
Based on the computation of the charge fraction for ions, these tendencies indicate that RPs in the region are more cooled and more strongly overionized, and can be naturally explained by a rapid cooling via thermal conduction.
Given the similar result for W44 reported by \cite{Okon2020}, evaporation of clumpy gas embedded in the hot plasma may cause the rapid cooling.
We have also discussed the possibility that ionization of protons accelerated in IC~443 is responsible for the overionization.
Based on the energetics of particle acceleration, we conclude that proton bombardment is difficult to explain the observed properties of the RPs.

\acknowledgments
We are grateful Dr. H. Sano for providing us with the NANTEN2 data used in the paper.
We appreciate all the {\it XMM-Newton}, NANTEN2, and IRSF team members. 
This work is partially supported by JSPS/MEXT Scientific Research grant Nos. JP19J14025 (H.O.), JP19H01936 (T.T.), JP19K03915 (H.U.), JP15H02090 (T.G.T.), JP19H00702 (M.S.), and NASA Grant 80NSSC18K0409 (R.K.S.).

\end{document}